# Automated Floorplanning for Partially Reconfigurable Designs on Heterogenrous FPGAs

Pingakshya Goswami, Dinesh Bhatia
University of Texas at Dallas
{pingakshya.goswam, dinesh}@utdallas.edu

*Abstract*—**Floorplanning problem has been extensively explored for homogeneous FPGAs. Most modern FPGAs consist of heterogeneous resources in the form of configurable logic blocks, DSP blocks, BRAMs and more. Very little work has been done for heterogeneous FPGAs. In addition, features like partial reconfigurability allow on-the-fly changes to the executable design that can result in enhanced performance and very efficient utilization of resources. In this paper, we have designed a floorplanner for Partially Reconfigurable (PR) designs in FPGA that smartly decides one of the three proposed resource allocation schemes to floorplan a particular type of reconfigurable region. We also propose a White Space Detection algorithm for efficient management of white space inside an FPGA in order to reduce the area and the wire length. The floorplanner is demonstrated on Xilinx Virtex 5 and Artix 7 FPGA architectures and can be easily integrated with existing vendor-supplied Place and Route tools. The main objective of the floorplanner is to reduce the wire length, minimize wasted resources and the area. The performance of our floorplanner is evaluated using MCNC benchmarks. We have compared our proposed floorplanner with other previously published results reported in the literature. We observe a substantial improvement in the overall wire length as well as the execution time. Also, the floorplanner was integrated with vendor supplied place and route tools (Xilinx Vivado) to automate the floorplanning flow. The automation process was tested on a partially reconfigurable median filter used in image processing applications.**

## I. INTRODUCTION

FPGAs are becoming more and more complex as different types of resources like Configurable Logic Blocks(CLB), Block RAM(BRAM), Digital Signal Processing(DSP) Blocks, Hard Processors, Transceiver core, etc. are added to a single die. Mapping designs on such heterogeneous architectures is a difficult problem and automated floorplanning and placement tools can greatly help to design using FPGAs. Advances in manufacturing technologies have allowed *Partial Reconfiguration* of an FPGA die. *Partial Reconfiguration* (PR) allows on-the-fly changes to design during runtime. This added feature adds complexity to the design as a given locality on an FPGA no longer holds a designing statically. The configuration changes with time and designing of such systems are not well supported by EDA tools. Although PR is one of the special features of FPGAs with lots of applications, this feature needs a significant amount of manual interference and involves a number of steps [1,2]. One of the biggest challenges in Partially Reconfigurable design flow is the floorplanning of the reconfigurable regions(RR) present in the design. The current EDA tools are not smart enough to automatically partition the PR design into static and reconfigurable regions and then floorplan the RRs. Hence, floorplanning has to be done manually for all the vendor supplied tools [2, 3]. Due to these reasons, the designer needs to have a deep understanding of the FPGA architecture which can vary from one FPGA device to another. Since most of these operations in partial reconfiguration have to be performed manually, especially the floorplanning part, the designs are not well optimized in terms of area, wire-length, resource wastage, and frequency of operation.

In this paper, we present the design of a floorplanner for partially reconfigurable designs in heterogeneous FPGAs that takes into consideration the variety of resources present in the FPGA device. The floorplanner is based on *fixed outline simulated annealing* algorithm. We propose a Priority-based algorithm mimicking *Olympic Medal Tally* for initial floorplanning. This preprocessing step is followed by simulated annealing based optimization. Also, a *White Space Detection* and storing algorithm for floorplanning of smaller module has been introduced. We defined a cost function, which is composed of a weighted sum of wire-length, area, and resource wastage. We consider the base architecture of the FPGA as Virtex 5 (for evaluating the performance of the tool) and the latest Artix 7 in which one real sample design has been floorplanned, implemented and configured on FPGA. Given that most of the Xilinx devices have similar architecture, our proposed methodology should easily map on to most of the devices in the Xilinx family. We integrated our software with Xilinx Vivado [8] to generate the floorplan of a partially reconfigurable *median filter* used in image processing applications. This application consists of seven reconfigurable regions and we have compared the manually generated floorplan with the one generated by the automatic floorplanner.

The rest of the paper is organized as follows: In Section II we discuss the related work. In Section III, we describe the problem and definitions of the terms used; the proposed floorplanner *PR_FP_tool* has been described in Section IV. Section V presents the experimental results. Finally, we conclude the paper and suggest some future extensions in Section VI.

## II. RELATED WORK

The problem of floorplanning in FPGA has been studied in the literature and, most of these works have addressed the floorplanning problem for static as well as partially reconfigurable designs on both uniform [7, 12 ,13] and nonuniform heterogeneous FPGAs [4,5,6]. Here, non-uniform refers to FPGAs that have uneven and scattered distribution of resources. The first available work for floorplanning of heterogeneous FPGAs was presented in [12] where the authors have targeted the floorplanning problem using a slicing tree based method and a modified version of Stockmeyer [16] floorplan optimization algorithm. This algorithm works well on the older generation of FPGAs like Spartan 3 and Virtex 2 devices from Xilinx. These older generation of devices make use of only one type of resources, namely the configurable logic block (CLB). In [13], the authors described a floorplanning method on Spartan 3 FPGAs that is based on fixed outline Parquet [15] floorplanner. In [4], the authors have defined a floorplanner for the Virtex 5 FPGA architecture and the floorplanning problem is represented as a sequence pair. The floorplan is optimized using simulated annealing and has been tested using MCNC benchmarks. The cost function defined in [4] is biased towards wirelength only. Also, the floorplanner does plan or account for wasted resources. A mixed integer linear programming (MILP) based FPGA floorplanner is proposed in [5] that takes into account most of the essential factors in the cost function which includes wirelength, resource wastage, and the perimeter of a device. The main drawback of this floorplanner is that it takes a long time to solve even for small design. This floorplanner gives a tight result but a lot of dead space may be generated which in turn increases the area and wirelength of the overall design.

The floorplanning methodology we proposed here focusses on shortcomings of previous works along with novel innovations. These include white space detection for insertion of black box modules, introduction of wasted resource, and area as parameters in the cost function and absolute automation of the process by integrating with Xilinx Vivado Design Suite. Our floorplanner gives a feasible solution in each and every move of optimization and there never arises a situation when the number of resources allotted for a Reconfigurable Region is less than the required number of resources.

## III. FLOORPLANNER AND FPGA ARCHITECTURE

The input to our *PR_FP_tool* tool is a synthesized netlist which represents the partially reconfigurable design in terms of resource requirement *i.e.* the number of CLBs, BRAMs and DSPs required by each module and the connectivity between each module is represented as a hypergraph netlist. Apart from the netlist, other input to the tool is the FPGA device architecture which is represented as an MxN matrix, where M is the number of rows and N is the number of columns in the device. The output of the tool gives the coordinates of each of the Reconfigurable Regions inside the FPGA fabric Figure 1 illustrates the tool with input and outputs type.

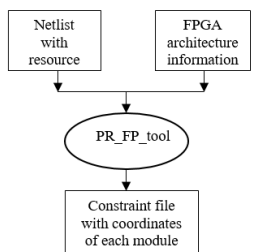

Fig. 1: *Inputs and Outputs of the proposed floorplanner.*

### A. Modelling of the FPGA Architecture

The FPGA architecture used for mapping is similar to the Xilinx Virtex 5 family of devices. Figure 2 illustrates an abstract view of one of the Virtex devices. The device has eight *clock-regions* laid out in four rows and two columns. A *clock-region* is 20 CLBs tall and spans half the length of the die. Each clock region consists of one of more number of DSP and BRAM columns. Majority of clock region is occupied by the CLBs. In the literature, the height of the clock regions inside the FPGA is considered as the size of the device used for mapping [5, 6, 9]. Some of the important architecture related definitions include:

*Static Region (SR)***:** Area inside the FPGA chip which remains constant during reconfiguration. The static region contains the processor or the control unit which decides when to switch between modules for the reconfiguration regions along with the clock and the I/O blocks.

*Reconfigurable Region/Reconfigurable Area (RR)*: Partitions inside the FPGA chip inside which the reconfigurable components reside. Reconfigurable regions can accommodate one or multiple instances of reconfigurable modules.

*Reconfigurable Modules(RM)*: Different instances of a functional unit which resides inside the RR. A RR accommodates different instances of RM keeping the input output pins similar. For example, A RR named "math" may have two instances of RM named adder and multiplier having the same pin connections.

*Frame*: The smallest reconfigurable unit present in a FPGA is called frame. A frame is one CLB column wide and one device row high [10]. Fig. 2 illustrates the structure of frames of different resources in Xilinx Virtex 5 FPGA device. A BRAM frame consists of 2 BRAM slices, DSP consists of 4 slices while a CLB frame consists of 20 CLB slices in Virtex 5.

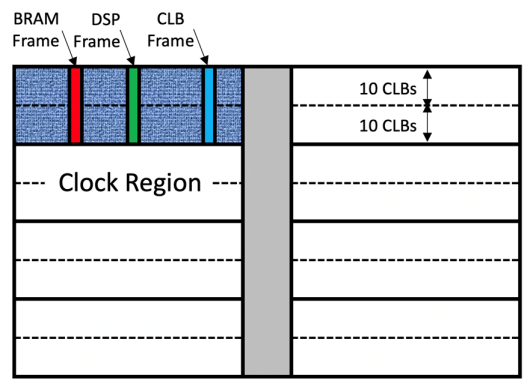

Fig. 2: *A Virtex 5 style FPGA architecture used for floorplanning. Each CLB frame is one CLB wide and has height that is same as the clock-region height.*

### B. Floorplanning Constrainsts for PR Design

In order to generate feasible floorplan, the following constraints are satisfied by our floorplanner *PR_FP_tool* in a heterogeneous FPGA:

i. None of the RRs should overlap. For two modules A and B not to overlap with each other, the following conditions are need to be satisfied:

$$(A_{x-min} \geq B_{x-max} \;||\; A_{x-max} \leq B_{x-min}) \;\&\&\; (A_{y-min} \geq B_{y-max} \;||\; A_{y-max} \leq B_{y-min}) \quad (1)$$

Here, *A* and *B* are two modules and *x-min* and *x-max* represents the extremes in horizontal direction while *y-min* and *y-max* represents the extremes in vertical direction of the two modules.

ii. Total resource available inside the FPGA chip must be greater than the sum of the resources of all RR and the SR

$$SR_{CLB} + \sum_{i=1}^{n} RR_{i_{CLB}} \leq FPGA_{CLB} \quad (2)$$
$$SR_{BRAM} + \sum_{i=1}^{n} RR_{i_{BRAM}} \leq FPGA_{BRAM} \quad (3)$$
$$SR_{DSP} + \sum_{i=1}^{n} RR_{i_{DSP}} \leq FPGA_{DSP} \quad (4)$$

In the above equations, SR represents resource consumned by static region and RR represents resource consumed bt reconfigurable regions. The total resource of each type inside the FPGA is represented by $PGA_{CLB}$, $FPGA_{BRAM}$ and $FPGA_{DSP}$ respectively.

iii. The rectangular area alloted to a particular RR must contain all the resources of each type required by the largest instance of PR module going into that region

$$RRa_{CLB} \geq max(RMa1_{clb}, RMa2_{clb} \ldots \ldots RMan_{clb}) \quad (5)$$
$$RRa_{BRAM} \geq max(RMa1_{BRAM}, RMa2_{BRAM} \ldots RMan_{BRAM}) \quad (6)$$
$$RRa_{DSP} \geq max(RMa1_{DSP}, RMa2_{DSP} \ldots RMan_{DSP}) \quad (7)$$

Here, $RR_a$ represents reconfigurable region named *'a'* and $RM_{a1}$, $RM_{a2}$ … $RM_{an}$ represents different instances of reconfigurable modules which are going into reconfigurable region $RR_a$.

iv. The shape, size and location of each static module should remain same across different instances of PR.

## IV. PROPOSED FLOORPLANNER

The entire design methodology consists of 8 stages which starts with synsthesis of the design in Xilinx Vivado[3] and ends with configuring the device using Vivado Hardware Manager. The important stages of the proposed design flow are:

i. Synthesize the static module using Xilinx Vivado and generate a netlist.
ii. Synthesize each of the partially reconfigurabe module using Xilinx Vivado and generate a netlist.
iii. Import the generated static and partial reconfiguration(PR) netlists to Vivado,
iv. Collect resource requirement statistics and wire connection information from Vivado and generate file that can be accepted by our *PR_FP_Tool*.
v. Run *PR_FP_tool* and generate a constraint file with the coordinates of each RR module
vi. Using the xdc file generated by the *PR_FP_tool*, create and implement design runs in Vivado..
vii. Generate bitstream for each configuration of PR designs
viii. Configure device and store PR bitstreams in on-board memory.

### A. Priority Sort of Reconfigurable Regions

The RRs present in the design are classified into four different categories:
- Type1: #DSP>0, #BRAM>0, #CLB>0
- Type2: #DSP>0, #BRAM=0, #CLB>0
- Type3: #DSP=0, #BRAM>0, #CLB>0
- Type4: #DSP=0, #BRAM=0, #CLB>0

Type1 modules are floorplanned first, followed by Type 2, Type 3 and Type 4. The order in which the RRs are to be allotted are stored in a *sorted region list*. The resources, which are more readily available, are given lower priority (e.g. CLB) and the ones which are least available are more costly and are given higher priority (e.g. DSPs).

The *sorted region list* created above is used for building the initial floorplan that will be perturbed during simulated annealing.

### B. Allocation of Reconfigurable Regions

As mentioned in the previous section, the RR present in a design is classified into four different categories. For the floorplanning of the four different types of RR, three distinct types of resource allocation schemes are adopted. These three resource allocation schemes are discussed here:

B.1. Resource Allocation Scheme 1

This scheme is used to floorplan RRs that consist of all the three types of resources (CLB, BRAM and DSP). First we allocate the DSP blocks required by the RR as DSP has the highest priority and most expensive in Virtex 5. After DSP requirement of a RR is fulfilled, BRAM blocks are allotted in a similar manner. DSP and BRAM block allocation is followed by CLB allocation. CLB frames are also addressable individually and one CLB frame contains 20 CLB blocks. Fig. 3 shows how a Type 1 RR is floorplanned in the FPGA. As shown in Fig.3, we allocate block labeled *A* first which is DSP, followed by block *B* (BRAM) and finally *C*, which are the surrounding CLBs.

B.2. Resource Allocation Scheme 2

This method is used to pack RRs that consists of only two types of resources; either DSP and CLB or BRAM and CLB. While allotting the CLB blocks, it is ensured that the device rows occupied by the CLB blocks are same as the DSP or BRAM rows. If more CLB blocks are required, CLB blocks towards the left or the right of the higher priority column are allotted. As shown in Fig.3, we allocate block labeled *D (DSP)* first which is DSP, followed by block *E* which represents the surrounding CLBs.

### C. White Space Detection and Calculations

White Space represents the largest continuous rectangles of free space which are available inside the FPGA fabric after the allocation of Static Region, Type 1 and Type 2 RRs. The white space detection algorithm scans the FPGA from left to right and finds the largest free rectangles inside the device. Once all the white space rectangles are formed, the coordinates and resource availability information of each of the rectangles are stored in a list called *white space region list*. This list is later used for Resource Allocation Scheme 3.

C.1 Resource Allocation Scheme 3

The white space available inside the FPGA is used for resource allocation of Type 4 RR which contains only one type of resource i.e. CLBs. The Type 4 RR are allocated to white spaces based on a cost function associated with each white space region. For each region, we compute a cost,

$$cost = \alpha * Free_{dsp} + \beta * Free_{bram} + \gamma * Free_{clb} + \delta * dist\_centroid \quad (8)$$

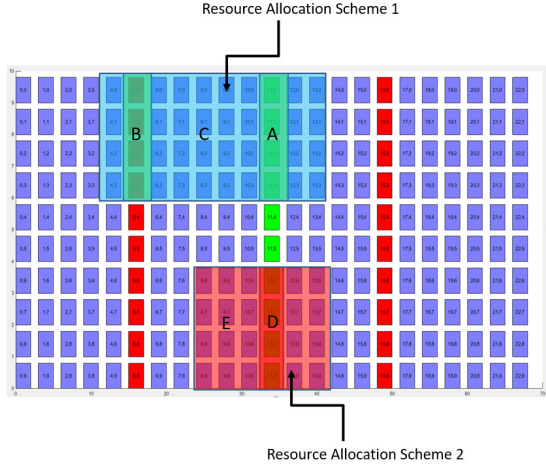

Fig. 3: *Illustration showing resource allocation scheme 1 and resource allocation scheme 2 inside the FPGA device.*

Here, $Free_{dsp}$, $Free_{bram}$ and $Free_{clb}$ represents number of free DSP, BRAM and CLB blocks respectively in the white space, and *dist_centroid* represents the distance of the white space centroid from the centroid of the whole design

$\alpha$, $\beta$, $\gamma$ and $\delta$ are weights assigned to each factor which can be customized by the user. We have considered $\alpha > \beta > \gamma > \delta$ as DSPs are more valuable as compared to other resources. The white spaces inside the white space region list are sorted in ascending order based on the cost provided by equation (8). In order to floorplan a Type 4 RR, we choose the white space which has the minimum cost as well as it contains sufficient number of CLB blocks to fulfill the CLB requirement of the RR. This scheme minimizes the wastage of high priority and costly resources. This list is recreated after allocation of each new RR inside the FPGA.

*D. Resource Wastage Calculation in an FPGA*

Owing to the fact that no two RR can share a same frame in the same *clock region* or in the same device row a number of resources are wasted which are external to the allotted rectangle of the RR. Here, we have shown the calculation for wasted resource calculation for each resource in a frame. A frame is one row high. We have assumed that the resources are allotted only in the vertical direction.

Definitions of terms used to calculate wasted resource in a row:
$n_{dsp}$, $n_{bram}$, $n_{clb}$ : number of DSP, BRAM, CLB blocks required respectively by a RR
$DSP_{per\,frame}$, $BRAM_{per\,frame}$, $CLB_{per\,frame}$ : number of DSP, BRAM and CLB blocks present in a frame respectively
$f_{dsp}$, $f_{bram}$, $f_{clb}$ : number of DSP, BRAM, CLBs frame required by a RR respectively

$RW_{dsp}$, $RW_{BRAM}$, $RW_{CLB}$ : number of DSP, BRAM, CLB blocks wasted respectively

Number of DSP blocks wasted is given by:
$$f_{dsp} = \left\lceil \frac{n_{dsp}}{DSP_{per-frame}} \right\rceil \quad (9)$$
$$RW_{dsp} = f_{dsp} * DSP_{per\_frame} - n_{dsp} \quad (10)$$

Number of BRAM blocks wasted is given by:
$$f_{bram} = \left\lceil \frac{n_{bram}}{BRAM_{per-frame}} \right\rceil \quad (11)$$
$$RW_{bram} = f_{bram} * BRAM_{per\_frame} - n_{bram} \quad (12)$$

Number of CLB blocks wasted is given by:
$$f_{clb} = \left\lceil \frac{n_{clb}}{CLB_{per-frame}} \right\rceil \quad (13)$$
$$RW_{clb} = f_{clb} * CLB_{per\_frame} - n_{clb} \quad (14)$$

*Observation*: If $n_{dsp}$, $n_{bram}$ and $n_{clb}$ are not integer multiples of $f_{dsp}$, $f_{bram}$ and $f_{clb}$ respectively, there will always be $RW_{rec}$ number of resources that are wasted. $RW_{rec}$ is given by:
$$RW_{rec} = f_{rec} * REC_{per\_frame} - n_{rec} \quad (15)$$
Here, *rec* represents the type of resource which can be CLB, BRAM or DSP. Resource wastage due to the restriction in sharing of a frame by two RRs occur if we allocate resource in the vertical direction. Since, CLBs are abundantly available in all directions, CLB block allocation in the horizontal direction in the same row should be given preference over extending CLB blocks in the vertical direction. This way, CLB resource wastage can be prevented significantly. But in case of BRAM or DSP, horizontal allocation of blocks are not possible as single block wide BRAM and DSP columns are aligned in the vertical direction only.

*E. Cost Function*

The initial floorplanning which is done via priority sorting algorithm and white space based region allocation method is perturbed by simulated annealing which minimizes the cost function given by equation (16).
$$cost = \alpha * WL + \beta * area + \gamma * WR \quad (16)$$
Here, *WL* represents half perimeter wirelength(HPWL), *area* represents total area of the design and *WR* total wasted resource in the design and $\alpha$, $\beta$ and $\gamma$ are weights assigned to each parameter by the user.

**Wirelength (WL)**: The wirelength is calculated using HPWL method. For a particular design, the total wirelength is calculated as the sum of HPWLs of all the nets present in the design.

**Area (Area):** The total area represents the area of the bounding box enclosing the entire design. The area is calculated as:
$$area = abs(X_{MAX} - X_{MIN}) * abs(Y_{MAX} - Y_{MIN}) \quad (17)$$
The *X* and *Y* coordinates of equation *(18)* represents the extremes of the entire design, not the reconfigurable regions.

**Wasted Resource (WR):** Wasted resource represents the blocks present inside a RR rectangle which are not actually utilized during placement. WR is calculated as:
$$rw_{cost} = \frac{total_{resource}}{total_{clb}} * clb_{wasted} + \frac{total_{resource}}{total_{bram}} * bram_{wasted} + \frac{total_{resource}}{total_{dsp}} * dsp_{wasted} \quad (18)$$

Equation (19) makes sure that the weight assigned to each wasted resource is proportional to the total number of elements of that particular resource present in the device. $rw_{cost}$ is passed to the cost function defined in equation (16).

## F. Representation of Floorplan and Simmulated Annealing

We represent the floorplan using sequence pair. The sequence pair consists of two sets of permutations of the blocks present in the design [19]. The two sequences present in a *sequence pair* depict the relative alignment of the modules with respect to each other. The following relationships represent sequence pair:
(<.....p,......q,....>,<.....p,......q,....>) => p is to the left of q
(<.....p,......q,....>,<.....q,......p,....>) => p is above q

Simulated annealing is used to minimize the cost function defined in equation (16) in order to generate an improved floorplan. The input to the annealer is the sequence pair of the current floorplan. The simulated annealing moves are described below:

**Move 0: Shuffle**: In this move, the annealer shuffles both the sequences in the sequence pair and starts floorplanning from the beginning. This move is very random and is generally performed at higher frequency when the temperature is high.

**Move 1: Swap two RR:** In this move, the annealer randomly chooses two RRs and swaps their positions in both the sequence pair. If two modules are swapped in both the sequences, they swap their positions in the actual floorplan.

**Move 2: Remove and Replace**: In this move, the annealer randomly chooses a RR and deletes it from both the sequences. The remaining RRs are floorplanned and the removed module is floorplanned after all others modules are packed.

**Move 3: Shift a RR Left/Right:** In this move, the annealer randomly chooses a RR and shift it either towards the right or to the left by certain number of CLB blocks. If the resource requirement of the RR is violated due the shifting operation, the move is ignored and the previous configuration is restored.

## V. EXPERIMENTS AND RESULTS

We have tested our floorplanner on an AMD Opteron(tm) Processor 3GHz work station. The floorplanner was tested on three different FPGA devices:
   a. A user defined 10x23 FPGA
   b. Xilinx Virtex 5 F110xt
   c. Xilinx Artix 7 XCA100T

The user defined 10x23 FPGA is a small heterogeneous FPGA which is modelled based on the Virtex architecture. It consists of 10 CLB rows and 23 columns. This will represent one *clock-region*. The columns are divided into 1 DSP column, 2 BRAM columns and remaining 20 CLB columns. The reconfigurable regions allotted to this FPGA are generated as a task flow graph using TGFF [14] graph generator. Also, it is assumed that the frame size is one, i.e. on each frame, only one type of resource will exist. Next, we have tested our floorplanner on Virtex 5 FPGA. We have used the 5 MCNC circuits and 1 GSRC circuit taken from [4] in order to evaluate the performance of our floorplanner. Finally, in order to validate our floorplanner we integrate it with Xilinx Vivado. We have taken the Xilinx Artix 7 device as our base architecture and floorplanned a partially reconfigurable median filter used in image processing applications using our tool. The median filter consists of seven reconfigurable regions.

## A. Result Analysis

We have evaluated the performance of our floorplanner using the 6 MCNC benchmarks. We calculated the wirelength, area, wasted resource and execution time for each of the benchmarks and compared our results against the existing work. The results presented here are based on Virtex 5 FPGA are shown in Table I and Table II.

TABLE I. WIRELENGTH COMPARISON WITH SIMILAR WORKS

|  | [13] | [7] | [12] | **Ours** |
|---|---|---|---|---|
| Circuit | HPWL | HPWL | HPWL | HPWL |
| apte | --- | 213540 | 2704 | 42940 |
| hp | --- | 113652 | 3286 | 40242 |
| xerox | --- | 536450 | 10476 | 112630 |
| ami33 | 89283 | 51356 | 4060 | 71280 |
| ami49 | 1173000 | 1001462 | 14050 | 218800 |
| n100 | 358338 | 132682 | 26355 | 189470 |

TABLE II. RUNTIME COMPARISON WITH SIMILAR WORKS

|  | [13] | [7] | [12] | **Ours** |
|---|---|---|---|---|
| Circuit | Time(sec) | Time(sec) | Time(sec) | Time(sec) |
| apte | - | 1.22 | 343 | 0.45 |
| hp | - | 0.96 | 531 | 0.47 |
| xerox | - | 1.02 | 353 | 0.58 |
| ami33 | 2.71 | 1.39 | 369 | 2.71 |
| ami49 | 4.95 | 3.84 | 585 | 4.59 |
| n100 | 8.86 | 8.87 | 573 | 15.24 |

As shown in Table I, we have calculated the wirelength as *half perimeter wirelength* (HPWL) which is in terms of CLB blocks. This means that the unit for wirelength is one CLB unit long. It has been found that our floorplanner *PR_FP_TOOL* gives an average 37% improvement of wirelength than the works mentioned by Banerjee *et. al*.[7]. In [7], the FPGA used for the testing the benchmark was Xilinx Spartan-3 XC3S5000 where the distribution of resources is uniform and follow a repetitive pattern. In this work, the authors have partitioned the FPGA into *basic tiles*. Each *basic til*e consists of 96 CLBs, 1 BRAM and 1 DSP block. The wirelength is measured in terms of *basic tiles*.

The work mentioned by Mehta *et. al.* in [13] was also tested on Spartan 3 and Virtex 2 FPGAs. Here the authors concentrated only on minimization of wirelength. Resource wastage is not accounted for in the cost function. The unit for calculation of wirelength is taken as CLBs which is similar to ours. When we compared our works with [13], we found that there is 49% improvement in wirelength for the last three benchmarks and the run time is comparable in both the cases.

The results in [12] were derived for Xilinx Spartan 3 XC3S500 FPGA. The floorplan was scaled [15] so that the bounding box of the entire floorplan is same as that of the FPGA chip. They have calculated the wirelength using center to center

HPWL of all the nets. The floorplanner takes much more time to generate the floorplan as compared to other state of the art works [4, 7, 13]. Our tool runs at around 500 times faster than the one mentioned in [12] since we apply pre-processing using *white space detection* and *priority based sorting* algorithms.

Table III shows the resource wastage for each type of resource. The average resource wastage for CLB, DSP and BRAM are 14.7%, 19% and 3.28% respectively which is very less as compared to other state of the art works [4,7,16].

TABLE III. PERCENTAGE OF RESOURCE WASTED FOR EACH CIRCUIT

| circuit | %clb wasted | %bram wasted | %dsp wasted |
|---|---|---|---|
| apte | 12 | 2 | 3 |
| hp | 12 | 5 | 0 |
| xerox | 11 | 2 | 3 |
| ami33 | 20 | 23 | 0 |
| ami49 | 17 | 21 | 14 |
| n100 | 17 | 40 | 3 |

### B. Integration With Vivado

We have integrated our floorplanner with Xilinx Vivado and tested on a filter designed for image processing application. The *PR_FP_TOOL* gives the output in a format which is similar to the *xdc* file which can be read by Xilinx Vivado tool. The image processing filter consists of 7 reconfigurable regions. Each reconfigurable region supports two reconfigurable modules:
    a. Median Filter
    b. Mean Filter

The resource requirement vector of actual design is significantly different from the resource requirement in MCNC benchmark. In the actual design, resource requirement is expressed in terms of SLICE_L and SLICE_M, RAM32 or RAM16 and DSP48 for Artix 7 FPGA. Figure 5(b) shows a design which is floorplanned by our tool on Vivado which is compared against a manually floorplaned design shown in Figure 5(a). Our floorplanner takes into account the edges to which the IOBs of the netlist are connected and places the modules near that edge.

When we compared the performance of our *PR_FP_TOOL* to a manual floorplanner, we found significant improvement in terms of area, maximum clock frequency and implementation time. It has been found that our floorplanner is able to reduce the total area of the reconfigurable modules by more than 80% and the frequency of the design has been increased by approximately 4%. Table IV compare the automatic floorplan performance against a manual one while Figure 5 shows the manual floorplan against the automatic one.

TABLE IV. MANUAL FLOORPLAN VS AUTOMATED FLOORPLAN

| Parameter | Manual | Automatic |
|---|---|---|
| Design Frequency | 26.85 MHz | 28.054MHz |
| Area | 3242 CLB$^2$ | 513CLB$^2$ |
| Place & Route Time | 4 hours | 19mins 23 secs |

## VI. CONCLUSION

We have presented a floorplanner for heterogeneous FPGAs that is capable of mapping partially reconfigurable designs. We have produced very efficient results with respect to the wirelength and the area. We have compared our work with the other approaches and have shown better results primarily due to the application of our novel white space detection algorithm that very effectively manages the dead space inside the device. Here, we have managed to reduce wasted resources by adopting three different resource allocation schemes for four different types of RRs. Also, our floorplanner can be easily integrated with vendor supplied place and route tools which we have proved by floorplanning a partially reconfigurable image processing median filter. In the future, this work can be extended to 3-D heterogeneous FPGAs.

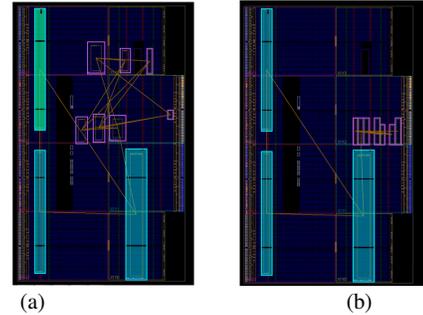

(a)            (b)

Fig. 5 *(a) Initial Manual Floorplan, (b) Optimized automatic floorplan generated by PR_FP_TOOL*